# TDACNN: Target-domain-free Domain Adaptation Convolutional Neural Network for Drift Compensation in Gas Sensors

Yuelin Zhang [a], Sihao Xiang [b], Zehuan Wang [a], Xiaoyan Peng [b,c], Yutong Tian [a], Shukai Duan [b,c], Jia Yan [b,c,*]

[a] WESTA College, Southwest University, Chongqing 400715, China

[b] College of Artificial Intelligence, Southwest University, Chongqing 400715, China

[c] Key Laboratory of Luminescence Analysis and Molecular Sensing (Southwest University), Ministry of Education, Chongqing 400715, China

[*] Corresponding author. E-mail address: yanjia119@swu.edu.cn (J. Yan).

## Abstract

Sensor drift is a long-existing unpredictable problem that deteriorates the performance of gaseous substance recognition, calling for an antidrift domain adaptation algorithm. However, the prerequisite for traditional methods to achieve fine results is to have data from both nondrift distributions (source domain) and drift distributions (target domain) for domain alignment, which is usually unrealistic and unachievable in real-life scenarios. To compensate for this, in this paper, deep learning based on a target-domain-free domain adaptation convolutional neural network (TDACNN) is proposed. The main concept is that CNNs extract not only the domain-specific features of samples but also the domain-invariant features underlying both the source and target domains. Making full use of these various levels of embedding features can lead to comprehensive utilization of different levels of characteristics, thus achieving drift compensation by the extracted intermediate features between two domains. In the TDACNN, a flexible multibranch backbone with a multiclassifier structure is proposed under the guidance of bionics, which utilizes multiple embedding features comprehensively without involving target domain data during training. A classifier ensemble method based on maximum mean discrepancy (MMD) is proposed to evaluate all the classifiers jointly based on the credibility of the pseudolabel. To optimize network training, an additive angular margin softmax loss with parameter dynamic adjustment is utilized. Experiments on two drift datasets under different settings demonstrate the superiority of TDACNN compared with several state-of-the-art methods.

## 1. Introduction

Electronic noses (E-noses), or artificial olfactory systems, are a series of cross-reactive gas sensors coupled with a set of pattern recognition algorithms that can simulate the mammal olfactory system to evaluate specific gas samples. An E-nose in the general sense is composed of three parts: a sensor array, a data acquisition and preprocessing apparatus, and a pattern recognition module.

Benefitting from its sensitivity to a wide variety of chemical compounds and complex mixtures and its qualitative/quantitative analysis ability, E-noses can achieve accurate recognition of complex samples, making them increasingly vital in food safety [1], environmental pollution monitoring [2], biomedical research [3], medical treatment [4], etc. However, due to the long-existing problems in the gas sensor manufacturing process, avoiding the unpredictable and nonlinear sensor drift phenomenon that occurs over a long period of time is still unrealizable [5, 6]. In addition, exposure to the same kind of analyte at a relatively high concentration leads to sensor poisoning, which can also result in sensor drift. Sensor drift deteriorates the gas sensor accuracy, creating issues for many aspects of real-life applications, such as research and industry, so drift compensation and domain adaptation algorithms for E-noses are indispensable.

The traditional domain adaptation methods for preventing gas sensor performance from deteriorating include 2 categories. One category carries out data correction on the data collected from the drift E-nose sensors, eliminating sensor characteristic deterioration. Yu et al. developed a data correction method based on principal component analysis (PCA) to correct deviations in the data. The results proved that using PCA to prealign the data can enable the pattern recognition algorithm to achieve better classification accuracy [7]. Independent component analysis (ICA) is another commonly used algorithm in data correction and is used to obtain mutually unrelated data components and remove drift disturbances [8]. However, since E-nose drift is nonlinear and unstable, data correction methods usually cannot deal well with the drift problem of E-noses. To compensate for this problem, a kernel transformation method was proposed by Tao et al. to perform the domain correction operation, which can improve the distribution consistency between the source domain and target domain [9]. Another is transfer learning, which projects the source domain data to the target domain space or projects the two domains' data to another regenerated space together. Transfer learning can reduce the distribution difference between the source domain and target domain so that the classifier trained with source domain data can have similar performance when predicting the data of the target domain. Zhang et al. proposed a domain regularized component analysis method (DRCA) to project the data before and after drift, giving the two subspaces after projection more similar distributions [5]. Zhang et al. also proposed a cross-domain discriminant subspace learning method (CDSL), which realizes drift compensation under the premise of ensuring data integrity [10]. As one of the improvement methods of CDSL, a local manifold embedding cross-domain subspace learning (LME-CDSL) algorithm is proposed by Tian et al., achieving subspace learning combined with domain adaptation and manifold learning [11]. For the extreme learning machine (ELM) algorithm, Wang et al. proposed an extreme learning machine with discriminative domain reconstruction, which can reduce the distribution difference between the source domain and target domain by learning a domain-invariant space [12].

In addition to traditional methods, although deep learning algorithms require considerably more computations than traditional algorithms, this high computing overhead has gradually become acceptable under the continuous development of hardware with the simultaneous growth of computing power. Some of them have been applied to E-nose data classification. The commonly used classification method is the multilayer perceptron (MLP), which achieves relatively good results due to its simple and clear structure. Gamboa et al. performed experiments based on a series of deep learning methods and proved that a multibranch MLP model consisting of only fully connected (fc) layers can achieve relatively better performance using less training data [13]. In addition, the original data of the E-nose are one-dimensional sequences composed of continuously collected voltage values. For the classification task of this sequence modeling problem, a recurrent neural network (RNN) or long short-term memory (LSTM) can be used. Julio et al. proposed an LSTM-based method and achieved dynamic data recognition from a metal oxide (MOX) gas sensor array using only 30 s of time sequences [14]. In [15], an LSTM model was introduced, achieving a good result in gas mixture concentration identification.

Additionally, various E-nose domain adaptation methods based on deep learning algorithms have been proposed and have shown good performance. Liu et al. proposed an autoencoder method that can automatically extract features from the original sensor sequence and perform classification under sensor drift [16]. Luo et al. applied a deep belief network (DBN) as a data preprocessing method based on unsupervised training; then, by cascading an SVM classifier, it successfully achieved drift compensation [17]. Tian et al. developed a method based on DBN and divided the operation into unsupervised and supervised parts to better recognize odor samples and compensate for sensor drift at the decision level [18]. In terms of LSTM, Wu et al. used LSTM as a dynamic feature extraction method to minimize feature dimensions and achieve E-nose drift compensation [19].

As an emerging subdiscipline of deep learning, convolutional neural networks (CNNs) have thrived in the field of image recognition in recent years due to their excellent feature extraction capabilities and generalization abilities. Many well-performing CNN models, such as LeNet-5 [20], AlexNet [21], Visual Geometry Group (VGG)-Net [22], ResNet [23], and InceptionNet [24], have emerged. Additionally, an increasing number of E-nose studies are focusing on CNNs. Many researchers attempt to use CNNs to preprocess E-nose data to improve cascaded classifier performance. A data preprocessing method was proposed by Wu et al., which can dynamically calculate the gradient of the original curve and perform sampling when the gradient is relatively large to reduce the size of the CNN input [25]. As one of the most classic CNNs, LeNet-5 and its modifications have been used in E-nose data pattern recognition and have achieved relatively good results, proving the effectiveness of CNN on E-nose sample classification tasks [26, 27]. However, existing attempts have failed to make targeted modifications to the E-nose features based on the original network structure of LeNet-5, and no relevant attempts have been made for domain adaptation. To better adapt CNNs to the data characteristics of E-noses (usually a one-dimensional time series), many studies have begun to focus on the customized modification of CNNs. Wang et al. proposed a CNN with a one-dimensional band convolution kernel for electronic nose data recognition and classification, achieving better results than traditional methods [28]. For mixture gas classification, a multilabel method based on a 1D CNN was proposed by Zhao et al. to identify multiple gas elements simultaneously in a complex odor mixture [29]. In terms of dynamic gas recognition, Qi et al. proposed a CNN-based gas dynamic identification method, achieving 95.7% accuracy when the sampling time reaches 15 s [30]. In addition to proposing new network structures, researchers have also drawn from the existing excellent CNN model in the computer vision field. Han et al. carried out a series of experiments to determine whether popular CNN structures used in computer vision (ResNet18, ResNet34, ResNet50, VGG-16 and VGG-19) can achieve satisfactory performance in mixed gas classification tasks [31]. Peng et al. designed GasNet with a shortcut structure for E-nose data classification and achieved good results [32].

In terms of domain adaptation methods based on CNNs, many have also been proposed and achieved good results in different fields [33, 34]. However, in E-nose drift compensation, CNN-based domain adaptation methods are still lacking. In addition, the existing deep learning domain adaptation methods achieved relatively good results on various public datasets under the prerequisite of obtaining sufficient uniformly distributed target domain data to help learn the target domain distribution and achieve domain alignment. The required target domain data usually require full-category samples, and some active learning methods even need a small batch of target domain samples associated with true labels to enhance performance [35, 36], which is difficult to fulfill in real scenarios. To the best of our knowledge, a domain adaptation method based on a CNN utilized in E-nose drift compensation has never been proposed.

In this work, we propose a CNN-based domain adaptation method called the target-domain-free domain adaptation convolutional neural network (TDACNN). Compared with the existing general algorithms, the proposed TDACNN introduces several innovations in the design of the network structure, and is able to complete the modeling and achieve higher accuracy without the participation of target domain data. It makes full use of the embedding features derived from each CNN level by using a multibranch and multiclassifier structure with a classifier ensemble method. It achieves acceptable results on two public drift E-nose datasets. In summary, the main contributions of this work are listed below:

(1) Different from most existing general algorithms, the proposed method only uses the source domain data without involving samples from the target domain, which can be more realistic and pragmatic in real-life applications where a large quantity of data from the target domain is usually unavailable.

(2) A novel multiclassifier end-to-end CNN is proposed to perform pattern recognition and domain alignment at the decision level, which does not require additional classifiers at the model's output end and has a simple structure for easy operation. The proposed classifier ensemble method based on maximum mean discrepancy (MMD) is adopted to ensemble the output of multiple classifiers. Also, in TDACNN, only a few parameters in loss function need to be manually set, which significantly reduces the difficulty of deployment.

(3) The proposed method works well with or without data preprocessing and feature preextraction, which can reduce the deployment difficulty while ensuring high accuracy. Failure to use the appropriate methods can be catastrophic to the performance of the model, which TDACNN is designed to avoid.

(4) TDACNN introduces an additive angular margin softmax loss with parameter dynamic adjustment to help train the network, which can optimize the geodesic distance margin directly and help the network map the features into space that has a larger interclass and smaller intraclass distance. The dynamic parameter adjustment method can enhance the robustness of the training procedure.

(5) Gaining experience from the mammalian olfactory system, in which signals from the same types of odorant receptor cells converge into the same olfactory bulb for further processing, the upper layers of TDACNN are designed to have multiple branches under the guidance of bionics. It can achieve highly customized feature extraction for different types of sensors. The multibranch structure gives the network flexibility to adapt to data in different forms and can be quickly deployed in application scenarios with different numbers of sensors simply by increasing or decreasing the number of branches.

## 2. Methodology

The structure diagram of the proposed TDACNN is shown in Fig. 1. After the batch normalization operation, the data from different sensors will be distributed to the corresponding branches to achieve customized feature extraction and drift compensation. The subsequent part of the network includes a multibranch structure for extracting features of each sensor separately, each of which contains three convolutional blocks for different levels of feature extraction and concentration. After each convolutional block, the path leading out from the branch connects to the classifier after width concatenation. These four classifiers can make full use of the results of each CNN level and obtain results based on the inherent characteristics of different levels. After the four classifiers give predictions, the TDACNN introduces a method called the MMD-based classifier ensemble, combining the output of all classifiers by jointly measuring the reliability of every pseudolabel. During training, an additive angular margin softmax loss in association with a parameter dynamic adjustment method is introduced to optimize the network training.

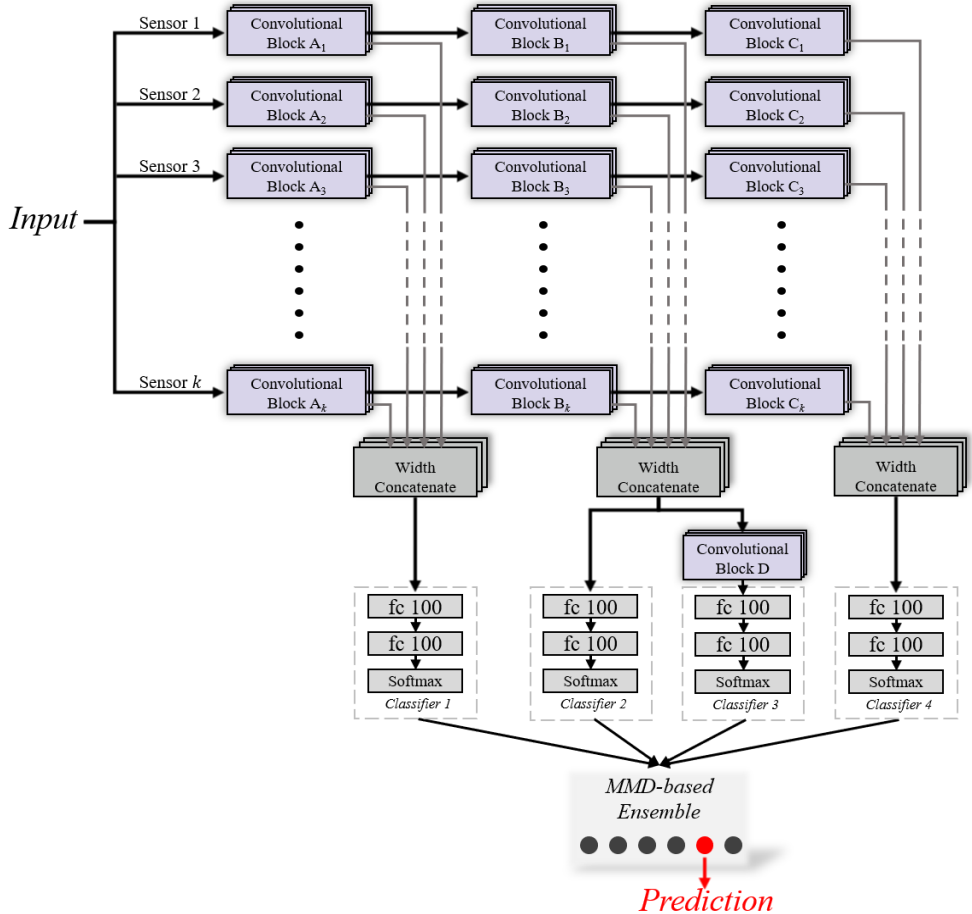

Figure 1. Structure of the proposed TDACNN.

## 2.1 CNN Structure Overview

2.1.1 Multibranch structure with one-dimensional convolution

The existing artificial olfactory system for real-time gas detection usually uses several sensors simultaneously. Due to this multiple gas sensor array solution, the raw data obtained by the system contain multiple dimensions composed of data collected from different sensors with variant response characteristics, which requires the utilization of customized feature extraction structures for different sensors. Therefore, a mutually independent multibranch structure is proposed to extract features from different sensors separately. The network shown in Figure 1 contains $k$ sensors; each sensor corresponds to a branch, so $k$ branches are presented. Another advantage of the multibranch design is that the network can adapt to different artificial olfactory systems more flexibly. For systems with different sensor configurations, TDACNN only needs to increase or decrease the number of branches to make adaptive changes, and then the network can learn the sensor variant feature automatically.

In the multibranch structure, 3 types of convolutional blocks are involved (convolutional blocks A, B and C). The convolutional blocks are composed of multiple layers, such as convolutional layers, max-pooling layers, or batch normalization layers. Each sensor dimension uses a set of independent convolutional blocks A, B and C, so each convolutional block can be trained to be a highly customized feature extractor suitable for different sensors, achieving better feature extraction that varies from sensor to sensor. In addition, the utilization of this hierarchical multibranch structure can also increase the width and depth of the network so that CNN can simultaneously acquire various features with richer levels, creating possibilities for subsequent operations.

2.1.2 Width concatenate

The tensor copies provided by the multibranch structure need to be concatenated to combine multiple feature maps into a whole before entering the classifier. Different from the depth concatenation (filter concatenation) performed by other multiclassifier networks such as InceptionNet [23], TDACNN uses width concatenation to splice multiple feature maps. This design adapts to the characteristics of E-nose data; that is, its data are usually composed of multiple one-dimensional time series collected by multiple sensors. Using concatenate width instead of concatenate depth can increase the width of each submap without increasing the number of submaps, thereby reducing the number of filters in the next level of the convolutional layer and leading to a considerable reduction in the number of parameters, which can reduce computing power overhead and avoid overfitting.

2.1.3 Multiple classifiers and two-dimensional convolution

As shown at the bottom of Fig. 1, four classifiers, namely classifier 1, classifier 2, classifier 2 and classifier 4, are connected to multibranch structures at different locations. This design learns domain-specific features and domain-invariant features at different CNN levels. All four classifiers have the same structure, which is two fc layers with 100 units cascaded with a softmax layer. Each fc layer performs a nonlinear mapping $\mathbf{h}_i^l = relu^l(\mathbf{W}^l \mathbf{h}_i^{l-1} + \mathbf{b}^l)$, where $\mathbf{h}_i^l$ and $\mathbf{h}_i^{l-1}$ are the output of the $i$-th node in the $l$-th fc layer and the upper fc layer, respectively, $\mathbf{W}^l$ and $\mathbf{b}^l$ denote the weights and bias of the $l$-th fc layer, and $relu^l$ represents the activation of the $l$-th layer and the rectifier units; $relu(\mathbf{x}) = \max(0, \mathbf{x})$ is utilized in all the fc layers in this proposed method. In the training phase, the losses contributed by all four classifiers are summed to the total network loss to help train and prevent gradient dispersion.

Regarding the location of the four classifiers, classifier 1 and classifier 2 are connected to the output of convolutional block A and convolutional block B, respectively, for classification based on the features derived from the shallower layers of the CNN. Classifier 3 is also connected to the output of the middle CNN layer. The difference is that classifier 3 is cascaded with convolutional block D to perform further feature extraction on the concatenated submaps. Here we introduce the fourth convolution block besides convolution blocks A, B, and C, i.e., convolution block D. It is worth noting that, unlike other convolutional blocks, the convolutional block D is located after the width concatenation. When designing convolutional block D, two-dimensional convolution is often used, which means convolutional block D extracts the intermediate feature and the interrelationship between different sensors, thereby widening the receptive field of the network. Classifier 4 is placed behind convolutional block C to process the high-level features derived after all three convolutional blocks on the branch. Since the features obtained by each classifier are different, the essence of this multiclassifier structure is to make full use of the embedding features from all CNN levels.

## 2.2 MMD-based Classifier Ensemble

There are 4 classifiers in TDACNN; however, the results of the four classifiers in the network cannot be considered simultaneously, and there is no way to know which one of results of the four classifiers is more reliable, resulting in a loss of useful information. To overcome this drawback, in this paper, the MMD-based classifier ensemble method is presented to comprehensively consider the results of all classifiers by evaluating the credibility of the pseudolabel using both MMD and the probability provided by the softmax layer in the classifier.

Starting from reviewing the concepts of MMD, as an effective nonparametric metric measuring the distance between two distributions, MMD has been widely used in many existing domain adaptation methods. Using the kernel function $\phi(\cdot)$, samples are mapped to a reproducing kernel Hilbert space (RKHS), and the MMD between source domain $s$ and target domain $t$ is defined as

$$\text{MMD}^2(s,t) = \sup_{\|\phi\|_{\mathcal{H}} \leq 1} \| E_{\mathbf{x}^s \sim s}[\phi(\mathbf{x}^s)] - E_{\mathbf{x}^t \sim t}[\phi(\mathbf{x}^t)] \|_{\mathcal{H}}^2, \tag{1}$$

where $\|\phi\|_{\mathcal{H}} \leq 1$ denotes a set of kernel functions in the unit sphere of RKHS $\mathcal{H}$, and $E_{\mathbf{x}^u \sim u}[\cdot]$ denotes the expectation of two distributions, $u \in \{s, t\}$. In this paper, $\phi(\cdot)$ refers to the radial basis function (RBF) kernel. For the convenience of calculation, the proposed method uses an empirical estimation form of MMD [37], which is given by

$$\text{MMD}^2(\mathcal{D}_s, \mathcal{D}_t) = \| \frac{1}{M} \sum_{i=1}^{M} \phi(\mathbf{x}_i^s) - \frac{1}{N} \sum_{j=1}^{N} \phi(\mathbf{x}_j^t) \|_{\mathcal{H}}^2, \tag{2}$$

where $\mathcal{D}_s = \{\mathbf{x}_i^s\}_{i=1}^M$ and $\mathcal{D}_t = \{\mathbf{x}_j^t\}_{j=1}^N$ denote two sets of samples that are drawn from source domain *s* and target domain *t*, respectively.

Different from MMD in some other domain adaptation methods, which involve complete access to the full-category target domain samples to learn the assumed target domain data distribution[5, 10, 11], MMD here refers to the discrepancy between the current test sample (target domain) and the labeled training samples (source domain), bringing more feasibility to real-life applications.

The workflow of the MMD-based classifier ensemble in a single classifier is shown in Figure 2. The MMD between the embedding features of the test sample $\mathbf{x}^t$ and source domain data is calculated based on the classification result of the classifier,

$$\mathrm{MMD}_n^2(\mathcal{D}_s^c, \mathbf{x}^t) = \|\frac{1}{M_c}\sum_{i=1}^{M_c}\phi(fc_n(\mathbf{x}_i^s)) - \phi(fc_n(\mathbf{x}^t))\|_{\mathcal{H}}^2, \quad (3)$$

where sample $\mathbf{x}^t$ is recognized as the *c*-th class by the *n*-th classifier, $\mathcal{D}_s^c$ denotes a set of samples drawn from the *c*-th class in the distribution *s*, and $M_c$ represents the number of the *c*-th class samples in $\mathcal{D}_s^c$. $fc_n$ denotes the embedding features derived from the second fc layer in the *n*-th classifier.

Derived from the softmax layer, the output of the classifier is in probability form, which indicates the probability that the tested sample belongs to a certain class. To ensemble MMD and the output of the classifier to obtain credibility, these two distributions need to be unified. We propose the standardized sigmoid as the activation function to remap the MMD into a unified distribution,

$$SS(x) = \frac{1}{1+e^{-x^*}}, \quad (4)$$

where $x^*$ indicates the zero means standardized version of *x*. After the activation, as the curve shows in Figure 2, the original MMD distribution is mapped into a distribution ranging from 0 to 1 to align with the distribution of the softmax output, which is named the activated MMD.

When predicting, there are two possible situations for the four classifiers. One of the situations is that all four classifiers derive the same result, and then the common result is considered the prediction result of the proposed method. However, the four classifiers usually derive several different results, so the different results should be evaluated jointly.

Using $y_n$ to represent the pseudolabel derived by the *n*-th classifier, $p_{y_n}$ refers to the probability of the $y_n$-th class provided by the *n*-th classifier's softmax layer. The ensembled credibility score of the *c*-th class ($ECS_c$) is presented as

$$\mathrm{ECS}_c = \sum_{y_n=c} \frac{p_{y_n=c}(\mathbf{x}^t)}{SS(\mathrm{MMD}_n^2(\mathcal{D}_s^c, \mathbf{x}^t))}. \quad (5)$$

After each WCS$_c$ is calculated, the class with the highest WCS$_c$ is considered the most credible class and is selected as the predicted label of TDACNN. The essence of MMD-based classifier ensembles is to consider the embedding features in several different stages in CNNs and evaluate which pseudolabel has a better domain adaptation outcome, i.e., which pseudolabel has the embedding feature with the shortest distance between the distribution of the source domain and target domain.

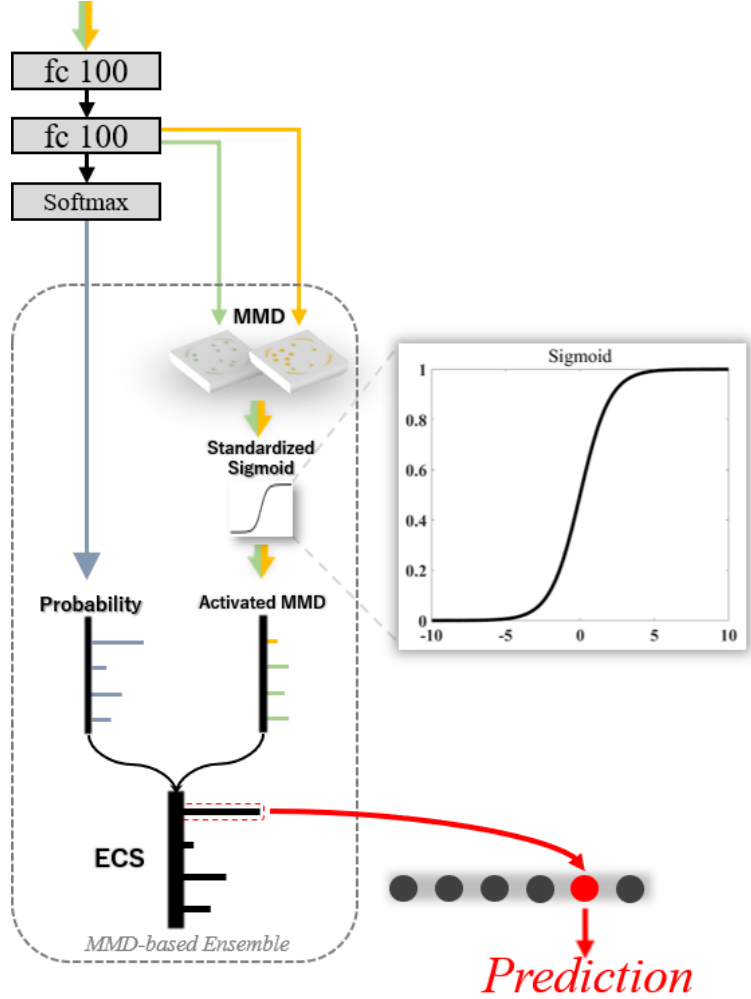

Figure 2. MMD-based classifier ensemble workflow.

## 2.3 Additive Angular Margin Softmax Loss

Inspired by [38], in this paper, an additive angular margin softmax loss with parameter dynamic adjustment is presented to further optimize the training. For softmax loss, there is a drawback that deteriorates the performance when faced with multiclass classification tasks; that is, the predicted result is not separable enough for the multiclass recognition problem leading to confusion between different samples, i.e., the traditional softmax loss does not explicitly optimize the features of samples from the same class and different classes, so that the trained network cannot respond correctly to samples from different classes. Additive angular margin softmax loss is proposed to improve the discriminative capability of the classifier in gas classification tasks. The basic idea is to calculate the angle between the target weight and the current feature after converting these two to angle form using the arc-cosine function and then add an additive angular margin to the angle. By performing this operation, the geodesic distance on the hypersphere between different classes can be increased, and the distance between samples from the same class can be reduced.

Start from the softmax loss,

$$L_1 = -\frac{1}{N}\sum_{i=1}^{N}\log\frac{e^{\mathbf{W}_c^T\mathbf{x}_i+\mathbf{b}_c}}{\sum_{j=1}^{C}e^{\mathbf{W}_j^T\mathbf{x}_i+\mathbf{b}_j}}, \qquad (6)$$

where the $j$-th column of the output weight $\mathbf{W}\in\mathbb{R}^{d\times C}$ is represented by $\mathbf{W}_j\in\mathbb{R}^d$ and $\mathbf{b}_j\in\mathbb{R}^C$ as the bias. $\mathbf{x}_i\in\mathbb{R}^d$ represents the output of the upper layer before the softmax layer corresponding to the $i$-th sample in the $c$-th class. $N$ is the batch size representation, and $C$ is the total number of classes, which is equal to the number of columns of $\mathbf{W}$. The dimension $d$ of the deep feature matrix is configured to 100 in all 4 classifiers in the proposed TDACNN. To facilitate calculation, the bias $\mathbf{b}_j$ is fixed to 0. Then, the logit is transformed as follows,

$$\mathbf{W}_j^T\mathbf{x}_i = \|\mathbf{W}_j\|\|\mathbf{x}_i\|\cos\theta_j, \qquad (7)$$

where $\theta_j$ is the angle between the embedding feature (i.e., the upper layer output) $\mathbf{x}_i$ and the weight $\mathbf{W}_j$. To make the classification only depend on the angle between feature $\mathbf{x}_i$ and weight $\mathbf{W}_j$ rather than the relative size of values, $\|\mathbf{W}_j\|$ is fixed to 1 by $l_2$ normalization. $\|\mathbf{x}_i\|$ is also fixed to 1 by the same method and then resscaled to $s$; thus, the embedding features obtained from the upper layer are distributed on a hypersphere with radius $s$. By using equation (7), the softmax loss is transformed into the following term:

$$L_2 = -\frac{1}{N}\sum_{i=1}^{N}\log\frac{e^{s\cos\theta_c}}{\sum_{j=1}^{C}e^{s\cos\theta_j}}. \qquad (8)$$

As the features are scattered around each class center on the hypersphere, to simultaneously optimize the compactness among samples from the same class and the discrepancy among the different class samples, an additive angular margin penalty $m$ is added to the angle $\theta_c$, which is in terms of the ground truth class,

$$L_3 = -\frac{1}{N}\sum_{i=1}^{N}\log\frac{e^{s(\cos(\theta_c+m))}}{e^{s(\cos(\theta_c+m))}+\sum_{j=1,j\neq c}^{C}e^{s\cos\theta_j}}. \qquad (9)$$

Regarding the two parameters $s$ and $m$ in this loss function, the scaler s indicates the extent to which the feature is scaled, and its change directly causes the loss to be unstable and to oscillate. Therefore, in the proposed TDACNN, the parameter $s$ is kept constant. For the additive angular margin penalty $m$, since $m$ denotes the penalty added to the angle, a larger $m$ leads to a larger loss. In the early training stage, the loss is at a relatively large value. At this time, adding penalty $m$ can increase the loss and optimize training. However, when training goes further, the loss gradually decreases and then stops decreasing and stays at a certain value, which is usually several orders of magnitude smaller than the value at the beginning of training. This means that if a larger penalty $m$ is added, a larger and nondecreasable loss will appear in the final training stage, resulting in a larger gradient when backpropagating, which causes the training process to oscillate and leads to unstable network performance.

To overcome this effect, a dynamic parameter adjustment method is introduced,

$$m(e) = m_0 - ve \quad (v \leq \frac{m_0}{e}), \tag{10}$$

where $e$ is the epoch, and $v$ denotes the changing rate. $m_0$ represents the initial value of $m$. It should be noted that $v$ must satisfy $v \leq \frac{m_0}{e}$ so that $m$ is always greater than zero. Using this method, $m$ can automatically decrease as the training progresses so that the loss at the end of training is smaller and the training process is more robust.

As a demonstration, we use the fixed $m$ and the proposed dynamic adjustment method to train the proposed CNN, and the obtained loss curves with respect to epochs are shown in Figure 3. It is obvious that at the later training stage, the loss stops at a fixed value if a fixed $m$ is used, leading to a large nondecreasable value when $m$ is large. However, when the dynamic adjustment method is adopted, $m$ will decrease automatically as the training progresses, leading to a smaller loss at the later training stage, which can enhance the training robustness.

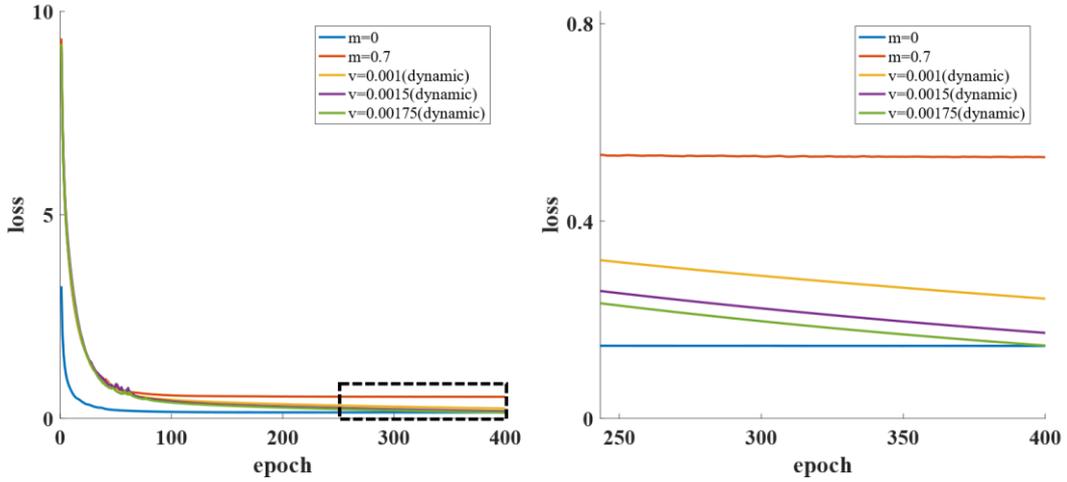

Figure 3. Loss curves with respect to epochs. The subfigure on the right is an enlarged version of the left subfigure from 250 to 400 epochs.

## 3. Experiment

### 3.1 Datasets

Two datasets are involved as benchmarks. The first dataset (Dataset A) is a public sensor drift dataset provided by the UC Irvine Machine Learning Repository [39]. The second dataset (Dataset B) mentioned in [40] was collected by researchers from Chongqing University.

3.1.1 Dataset A

This resource contains 13,910 samples from 16 E-nose sensors collected over 3 years, which are divided into 10 batches according to the time sequence to simulate sensor drift (Table 1). The dataset is composed of data from six distinct pure gaseous substances (ammonia, acetaldehyde, acetone, ethylene, ethanol, and toluene) at various concentrations. Each of the sensors contains 8 features, so a single sample can be represented as a vector of 128 dimensions (16×8). With a total size of 13910×128, this dataset is a large dataset in the field of E-noses. Though only a small portion of the samples in dataset A is used for training in each experiment, a larger total sample size allows for a relatively more considerable amount of data per batch. Thus, each experiment in dataset A can have more samples for training, which is beneficial to deep learning implementation. It can be seen in the scatter points in Fig. 4 that drift occurs throughout the whole period, and the distribution of the first two principal components varies significantly from the beginning to the end.

3.1.2 Dataset B

As shown in Table 2, dataset B contains 159 samples from 32 sensors collected over 4 months. It performs experiments on 7 kinds of gas at different concentrations obtained by evaporating beverages. Each sample in this dataset consists of a sequence with a length of 50, which is the original voltage response collected from the sensors. In the 1st, 3rd, and 4th months, 33, 63 and 63 samples were gathered, respectively. These data were divided into 3 batches according to the time sequence. As illustrated in Fig. 5, the distributions of batch 2 and batch 3 are similar because the acquisition time is closer. For the distribution of batch 1, since it is collected in the first month, the distribution demonstrates a considerable discrepancy compared with the last two batches.

Table 1. Details about dataset A

| Batch ID (Months ID) | Number of samples | | | | | | |
|---|---|---|---|---|---|---|---|
| | Ethanol | Ethylene | Ammonia | Acetaldehyde | Acetone | Toluene | Total |
| Batch 1 (Months 1,2) | 83 | 30 | 70 | 98 | 90 | 74 | 445 |
| Batch 2 (Months 3,4,8,9,10) | 100 | 109 | 532 | 334 | 164 | 5 | 1,244 |
| Batch 3 (Months 11,12,13) | 216 | 240 | 275 | 490 | 365 | 0 | 1,586 |
| Batch 4 (Months 14,15) | 12 | 30 | 12 | 43 | 64 | 0 | 161 |
| Batch 5 (Months 16) | 20 | 46 | 63 | 40 | 28 | 0 | 197 |
| Batch 6 (Months 17,18,19,20) | 110 | 29 | 606 | 574 | 514 | 467 | 2,300 |
| Batch 7 (Months 21) | 360 | 744 | 630 | 662 | 649 | 568 | 3,613 |
| Batch 8 (Months 22,23) | 40 | 33 | 143 | 30 | 30 | 18 | 294 |
| Batch 9 (Months 24,30) | 100 | 75 | 78 | 55 | 61 | 101 | 470 |
| Batch 10 (Months 36) | 600 | 600 | 600 | 600 | 600 | 600 | 3,600 |

Table 2. Details about dataset B

| Batch ID (months ID) | Number of samples | | | | | | | |
|---|---|---|---|---|---|---|---|---|
| | Beer | Black tea | Green tea | Liquor | Oolong tea | Pu'er tea | Wine | Total |
| Batch 1 (Months 1) | 3 | 6 | 6 | 3 | 6 | 6 | 3 | 33 |
| Batch 2 (Months 3) | 9 | 9 | 9 | 9 | 9 | 9 | 9 | 63 |
| Batch 3 (Months 4) | 9 | 9 | 9 | 9 | 9 | 9 | 9 | 63 |

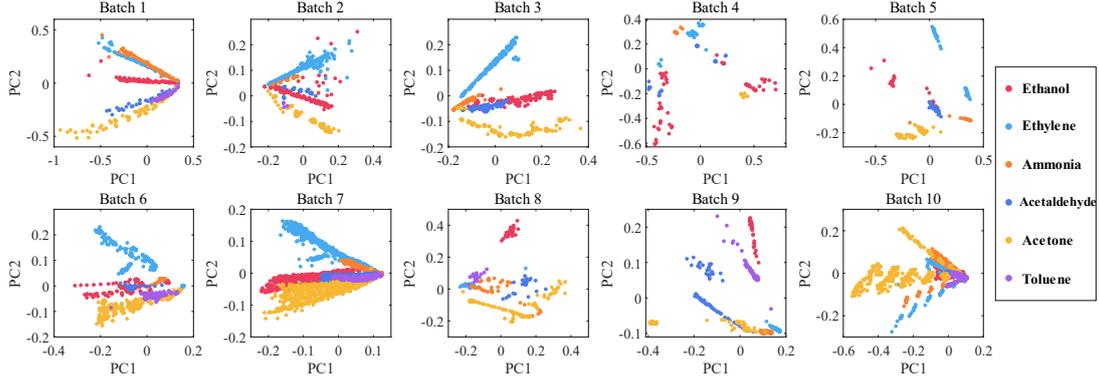

Figure 4. The PCA score plot of the first two principal components of the 10 batches in dataset A.

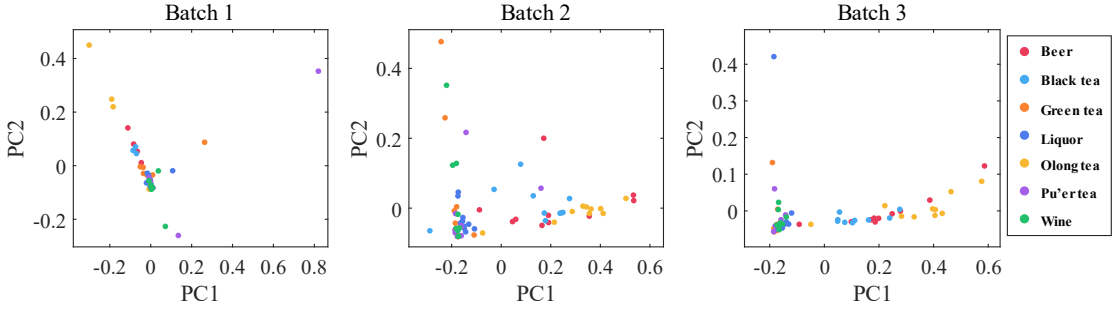

Figure 5. The PCA score plot of the first two principal components of the 3 batches in dataset B.

## 3.2 Experimental configuration

To simulate the long-term-drift scenario and short-term-drift scenario, we design 2 scenario settings, as shown in Table 3. For the long-term-drift scenario experiment, setting 1 uses the first batch as the source domain for model training and the remaining batches as the target domain for prediction. For the short-term-drift scenario, setting 2 is proposed to train the network using the batch before one batch and test the performance on this batch. Therefore, there are 9 experiments on setting 1 for both dataset A and dataset B and 9 and 2 experiments on setting 2 for dataset A and dataset B, respectively. Since the format and the characteristics of these two datasets are different, it is necessary to design appropriate convolution blocks.

As mentioned above, dataset A is a preprocessed and pre-extracted high-level dataset. This means that the data from dataset A do not require deep feature extraction using too many convolutional layers. The feature vector for each branch is a one-dimensional vector with a length of 8, which is too short for the proposed network. To compensate for these effects, for this dataset only, we design a relatively shallow convolutional block without a pooling layer to perform less feature extraction. As shown in the right column of Figure 6, there is a series of 1D convolutional layers with kernel sizes of $2 \times 1$ in convolutional blocks A, B and C. Three cascaded 2D convolutional layers with kernel sizes of $3 \times 3$ are utilized in convolutional block D to extract the interrelationship between different sensor dimensions. The activation function *ReLU* is utilized in the convolutional layer in this network.

For dataset B, since it contains the original time sequence, which needs more extraction and concentration than the data in dataset A, more convolutional layers are utilized. A 1D convolution is utilized in convolutional blocks A, B and C, and 2D convolution is utilized in convolutional block

D. As shown in the middle column of Figure 6, convolutional blocks for dataset B contain more convolutional layers and max-pooling layers than the blocks for dataset A. For this network, we use *ReLU* as the activation function for all convolutional layers.

Table 3. Experimental setting

| Scenarios | Data | |
|---|---|---|
| | Source domain | Target domain |
| Setting 1 | Batch 1 | Other batches |
| Setting 2 | Batch $i$ | Batch $(i+1)$ |

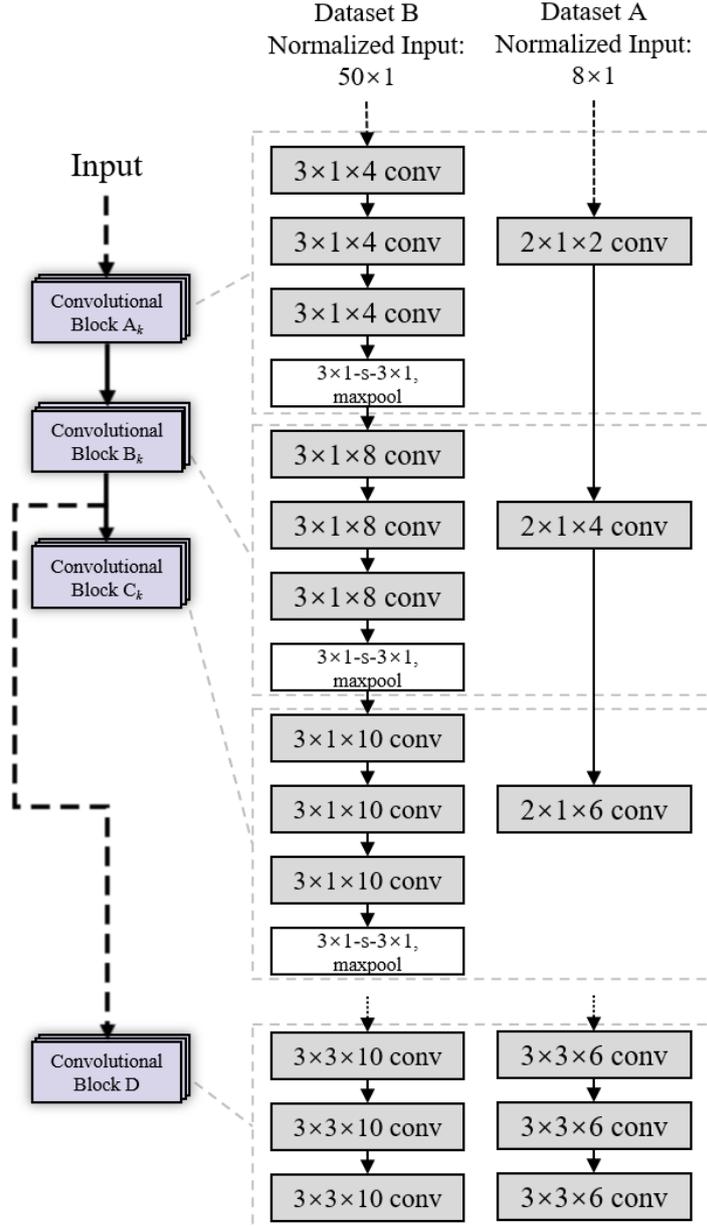

Figure 6. Structure of convolutional blocks. The left column is for dataset B, which has a normalized input map of 50×1. The right column is for dataset A with a normalized input map size of 8×1. In the convolutional layer (conv) in the figure, 3×3×8 means that the kernel size is 3×3, and the number of filters is 8. For the max-pooling layer (maxpool), 2×1-s-2×1 means that the kernel size for max pooling is 2×1 and the stride is 2×1.

## 3.3 Implementation details

To evaluate the performance of the proposed methodology, several state-of-the-art methods are introduced as comparisons. During selection, several guidelines are considered, including the concepts behind the methods, the similarity of structure, the size of the network, and the original application of the methods. Several classic CNN approaches are chosen, including the VGG series (VGG13, VGG16, VGG19) [21], AlexNet [20], LeNet-5 [19], ResNet34 [22], and GoogLeNet (InceptionV1) [23]. To be consistent with TDACNN, some CNNs with a large number of convolutional layers far beyond the number of convolutional layers in the proposed method are not considered. For CNNs in the E-nose field, several CNN-based methods utilized in artificial olfaction are also involved, including SniffMultinose [41], SniffConv [41], optimized deep CNN (ODCNN) [27], and GasNet [31]. In terms of methods based on general concepts except for CNN, we choose several methods, including PCA, linear discriminant analysis (LDA), locality preserving projections (LPP), (component correction based on PCA) CC-PCA [42], orthogonal signal correction (OSC) [43], generalized least squares weighting (GLSW) [44], and direct standardization (DS) [45]. An SVM with an RBF kernel (SVM-RBF) is cascaded to this general method as a classifier.

For all the CNN-based methods, we use Keras (using the TensorFlow backend) to build the network. During the training phase, we use stochastic gradient descent (SGD) as the optimizer and set the maximum training epochs to 400. To avoid overfitting, an early stopping strategy is applied to stop the training when the drop in loss is small enough. The proposed additive angular margin softmax loss with parameter dynamic adjustment is utilized in the TDACNN training. For the configuration of additive angular margin softmax loss, scaler $s$ is set to 4, the initial value $m_0$ of $m$ is set to 0.7, and the changing rate $v$ is set to 0.00175. For the general methods, we use MATLAB to perform the algorithm based on the corresponding works.

## 3.4 Experimental results

The experiments were conducted on Setting 1 and Setting 2, and the performance of different methods is reported in Table 4. For performance on dataset A, by observing the first two columns in Table 4, we found that the proposed TDACNN achieves the best performance in dataset A on both setting 1 and setting 2, which are 72.22 and 81.48, respectively. Among the general methods, LDA achieves relatively better results (66.48 and 69.64), but it is still lower than the accuracy of the TDACNN. The best performing method among the CNN-based counterparts is LeNet, which achieves better performance than that of the others in both setting 1 and setting 2, which are 61.84 and 69.18, respectively. The complete set of results of the TDACNN on dataset A is shown in Table 5. In terms of performance on dataset B, the TDACNN achieves the best performance on both setting 1 and setting 2, which are 63.49 and 69.05, respectively. SVM classifier without drift compensation and with PCA achieve the best result on setting 1 and setting 2 among the general methods, respectively. In the CNN-based method, the SniffMultinose achieves the best performance in both setting 1 and setting 2.

Table 4. Performance of different methods for dataset A and dataset B under setting 1 and setting 2.

| | Method | Classification accuracy (in %) | | | |
|---|---|---|---|---|---|
| | | Dataset A | | Dataset B | |
| | | Setting 1 | Setting 2 | Setting 1 | Setting 2 |
| CNN-based method | VGG19 | 21.02 | 20.81 | 14.29 | 14.29 |
| | VGG16 | 21.02 | 19.36 | 14.29 | 14.29 |
| | VGG13 | 21.02 | 37.15 | 14.29 | 14.29 |
| | AlexNet | 56.92 | 61.39 | 29.37 | 21.43 |
| | LeNet | **61.84** | **69.18** | 40.48 | 46.03 |
| | GoogLeNet | 22.46 | 65.51 | 35.71 | 43.65 |
| | Resnet34 | 43.31 | 63.66 | 31.75 | 42.86 |
| | SniffMultinose | 51.66 | 59.67 | **44.44** | **47.62** |
| | SniffConv | 23.86 | 54.97 | 18.25 | 41.27 |
| | ODCNN | 56.68 | 68.36 | 18.25 | 25.40 |
| | GasNet | 36.36 | 66.75 | 28.57 | 33.33 |
| General method | SVM | 63.55 | 68.29 | **38.89** | 43.65 |
| | LDA | **66.48** | **69.64** | 34.13 | 42.06 |
| | CC-PCA | 60.29 | 51.53 | 23.81 | 26.19 |
| | PCA | 63.81 | 68.90 | 38.10 | **45.24** |
| | LPP | 54.57 | 60.88 | 36.51 | 35.71 |
| | DS | 35.97 | 37.49 | 20.63 | 20.63 |
| | GLSW | 50.13 | 57.54 | 33.33 | 31.75 |
| | OSC | 59.68 | 53.10 | 28.57 | 24.60 |
| | **TDACNN** | **72.22** | **81.48** | **63.49** | **69.05** |

Table 5. Performance of the TDACNN trained on batches 1–9 and tested on successive batches (on dataset A).

| Batch ID | Classification accuracy (in %) on batches 2–10 | | | | | | | | |
|---|---|---|---|---|---|---|---|---|---|
| | 2 | 3 | 4 | 5 | 6 | 7 | 8 | 9 | 10 |
| batch 1 | 89.56 | 83.83 | 77.64 | 75.63 | 74.36 | 62.08 | 75.10 | 60.85 | 50.88 |
| batch 2 | | 97.46 | 86.60 | 84.77 | 76.23 | 73.71 | 56.80 | 62.34 | 52.25 |
| batch 3 | | | 87.58 | 75.13 | 76.01 | 74.43 | 52.04 | 71.70 | 52.86 |
| batch 4 | | | | 94.68 | 63.56 | 61.80 | 45.92 | 74.68 | 45.36 |
| batch 5 | | | | | 73.90 | 50.57 | 76.19 | 51.91 | 50.96 |
| batch 6 | | | | | | 80.18 | 53.74 | 61.28 | 59.82 |
| batch 7 | | | | | | | 78.43 | 60.85 | 65.84 |
| batch 8 | | | | | | | | 83.19 | 72.22 |
| batch 9 | | | | | | | | | 47.64 |

We provide the first two principal components after nonlinear projection by TDACNN to demonstrate the projection performance. Using the model trained under setting 1 with dataset A, the hidden layer (the second fc layer in classifier 2) output of ten batches in dataset A is derived, and the PCA scatter points are shown in Fig 7. These PCA scatter points can roughly show the sample distribution after the projection by TDACNN, from which the drift compensation performance of the network can be evaluated. It can be seen from Fig 7, after the projection by the proposed network, the distribution consistency between the drifted batch (batch 2 to batch 10) and the nondrifted batch (batch 1) is improved compared to Fig 4, which is without projection.

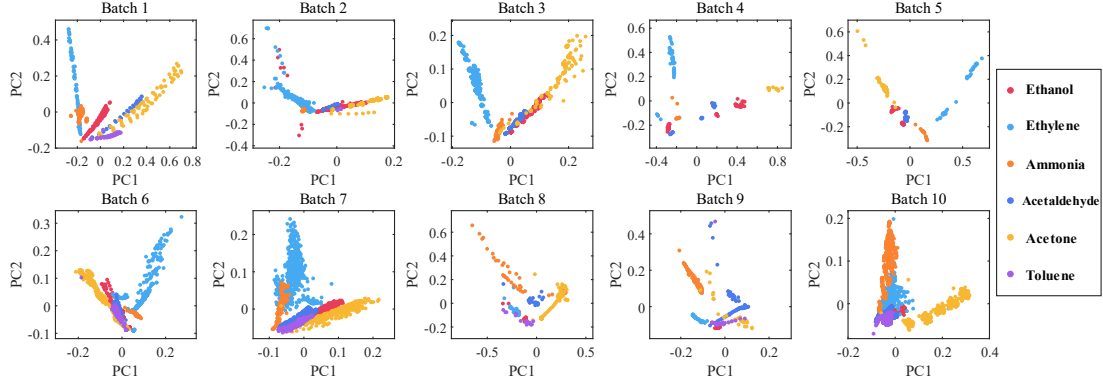

Figure 7. The PCA score plot with projection by the proposed TDACNN for the 10 batches in dataset A under setting 1.

## 3.5 Analysis of the MMD-based ensemble

In TDACNN, the MMD-based ensemble is utilized to combine the results of all 4 classifiers. This algorithm calculates the credibility of each pseudolabel and selects the most credible result. The performance of these 4 classifiers and the result of the MMD-based ensemble under setting 1 and setting 2 of dataset A and dataset B are shown in Tables 6 to 9. It can be observed in these tables that, in most of the cases, the performance of TDACNN after MMD-based ensemble can always be approximately equal to the performance of the best performing classifier, sometimes even better than the performance of each classifier. For example, in the experiment on 9→10 of dataset A under setting 2 (Table 7), the accuracy after ensemble reaches 47.64, which is higher than the best performing single classifier (46.11).

The average accuracy after the ensemble is higher than the other four classifiers' average accuracy in all 4 scenarios. Therefore, the ensemble effectiveness is proven.

Table 6. Performance (%) of different classifiers for dataset A under setting 1.

| Task | 1→2 | 1→3 | 1→4 | 1→5 | 1→6 | 1→7 | 1→8 | 1→9 | 1→10 | Average |
|---|---|---|---|---|---|---|---|---|---|---|
| MMD-based ensemble | 89.56 | 83.83 | 77.64 | 75.63 | 74.36 | 62.08 | 75.10 | 60.85 | 50.88 | 72.22 |
| Classifier 1 | 89.56 | 83.83 | 67.70 | 75.63 | 74.53 | 62.75 | 50.34 | 57.45 | 52.49 | 68.25 |
| Classifier 2 | 89.40 | 83.51 | 77.64 | 75.63 | 74.23 | 61.58 | 54.08 | 59.36 | 51.94 | 69.71 |
| Classifier 3 | 89.81 | 83.89 | 80.75 | 75.63 | 74.58 | 63.38 | 71.43 | 60.64 | 50.46 | 72.28 |
| Classifier 4 | 88.92 | 83.77 | 75.16 | 75.63 | 73.92 | 62.05 | 76.87 | 60.85 | 49.46 | 71.85 |

Table 7. Performance (%) of different classifiers for dataset A under setting 2.

| Task | 1→2 | 2→3 | 3→4 | 4→5 | 5→6 | 6→7 | 7→8 | 8→9 | 9→10 | Average |
|---|---|---|---|---|---|---|---|---|---|---|
| MMD-based ensemble | 90.29 | 97.46 | 87.58 | 94.68 | 73.90 | 80.18 | 78.43 | 83.19 | 47.64 | 81.48 |
| Classifier 1 | 90.05 | 96.58 | 82.61 | 96.45 | 74.14 | 73.43 | 76.87 | 82.77 | 34.22 | 78.57 |

| Classifier 2 | 90.29 | 95.81 | 84.47 | 75.13 | 60.43 | 80.18 | 75.51 | 64.89 | 38.40 | 73.90 |
| Classifier 3 | 90.29 | 94.04 | 85.09 | 83.76 | 70.61 | 78.66 | 78.57 | 83.19 | 46.11 | 78.93 |
| Classifier 4 | 90.45 | 98.80 | 87.58 | 74.62 | 58.64 | 70.16 | 77.89 | 73.40 | 39.84 | 74.60 |

Table 8. Performance (%) of different classifiers for dataset B under setting 1.

| Task | 1→2 | 1→3 | Average |
| --- | --- | --- | --- |
| MMD-based ensemble | 71.43 | 55.56 | 63.49 |
| Classifier 1 | 66.67 | 50.79 | 58.73 |
| Classifier 2 | 65.08 | 42.86 | 53.97 |
| Classifier 3 | 65.08 | 44.44 | 54.76 |
| Classifier 4 | 69.84 | 53.97 | 61.90 |

Table 9. Performance (%) of different classifiers for dataset B under setting 2.

| Task | 1→2 | 2→3 | Average |
| --- | --- | --- | --- |
| MMD-based ensemble | 71.43 | 66.67 | 69.05 |
| Classifier 1 | 66.67 | 66.67 | 66.67 |
| Classifier 2 | 65.08 | 46.03 | 55.56 |
| Classifier 3 | 65.08 | 41.27 | 53.17 |
| Classifier 4 | 69.84 | 41.27 | 55.56 |

## 3.6 Discussion

It can be seen in the experimental results that TDACNN achieves better results than all the methods used for comparison. Considering that the proposed method does not need to use target domain data in the training phase, the training and deployment procedure is no different from the general CNN used for classification, which can greatly reduce the operation difficulty and the model complexity. In addition, further analysis proved the effectiveness of MMD-based classifier ensembles. During the ensemble, the results with lower credibility are discarded, and the most reliable pseudolabel is successfully selected, leading to a better performance than any other individual classifier.

However, it is worth noting that the quantity of data in the batches in Dataset A is not uniform, leading to a deterioration in the performance to a certain extent. As presented in Table 1, for batches with a large quantity of data, such as batch 10, the number reached a considerable 3600. However, batches with fewer data, such as batch 4, contain only 161 data points, which is far too little for CNN training. When the CNN encounters a situation in which the data included in the training batch are much fewer than the data included in the validation batch, which leads to severe performance degradation due to the bias between the training samples and validation samples. For example, it can be observed in Table 7 in experiments 9→10. The quantity of data in batch 9 and batch 10 is 470 and 3600, respectively, and the accuracy after the ensemble is 47.64, which is the worst performance among all the experiments of dataset A under setting 2. In addition, for the experiments of dataset A under setting 1, batch 1 is utilized as the training set, which only has 445 samples.

However, for the target domain, there are 13465 samples in total, which is a large sample number bias between the source domain and target domain. The experimental results show that our method always obtains better results compared to the contrast methods when using the same training set, regardless of whether the training sample size is larger or smaller. This indicates that our method has better adaptability for datasets with different sizes compared to the contrast method.

Observing the performance of the other methods, LeNet is the best performing network for dataset A, which shows the effectiveness of the simple network structure. For dataset B, the best performing CNN among those CNNs is SniffMultinose, which achieves the highest accuracy in both setting 1 and setting 2, proving that the customized design of CNN structure for the E-nose domain adaptation and recognition task is necessary. For the general methods, LDA achieves better results on both setting 1 and setting 2 of dataset A since LDA modeling needs the source domain sample labels to be involved. In addition, among the general methods, 3 methods involve data from the target domain: DS, GLSW, and OSC. However, they are still at a lower level of accuracy when compared with TDACNN, although no samples from the target domain are involved during TDACNN's training or prediction.

Apart from the above methods we used for comparison, we also noted some other algorithms achieving even higher accuracy than the proposed method under similar situation, such as multi-feature kernel semi-supervised joint learning model (MFKS), domain regularized component analysis (DRCA), domain adaptation extreme learning machines (DAELM), transfer sample-based coupled task learning with standardization error-based model improvement (TCTL-SEMI), and active learning (AL) in [40]. However, we believe that they are not belong to the same class of method as the proposed TDACNN due to the different potential application scenarios and conditions. These methods modeling requires labeled and/or unlabeled samples from target domain while the TDACNN modeling does not require the participation of the target domain data. That means these methods achieve higher accuracy under the condition of relying on a large amount of target domain information, which is not met in TDACNN. Therefore, to enhance the representativeness and persuasiveness of the contrast experiments, most of the contrast methods reported in this paper are similar to TDACNN which only use source domain data.

Compared with those contrast methods, although the computational overhead of TDACNN is slightly greater, the advantages of the proposed method are significant. The most significant superiority is that TDACNN is able to accomplish drift compensation and pattern recognition without the involvement of target domain data. This advantage means that TDACNN is more practical for engineering purposes, and achieving higher accuracy without target domain data than models that require target domain data is a demonstration of the superiority of our proposed model. Second, in TDACNN, most of the parameters can be automatically trained by the back-propagation algorithm, only a few parameters need to be manually set to help training, and changing them will not cause the model structure to change. This means TDACNN can be deployed in the real-world situation under unconstrained and uncontrollable conditions and achieve higher accuracy without manual intervention and additional data. Third, TDACNN is a decision-level drift compensation method based on an end-to-end network, which only requires input data to obtain classification results and has a simple structure for easy operation. In addition, TDACNN achieves better accuracy with or without data preprocessing and feature extraction. But the prerequisite for the contrast methods to achieve reported accuracy is to use appropriate data preprocessing and feature extraction methods. This empirical operation increases the complexity of the algorithm.

However, the sensor will deteriorate during operation due to external factors such as sensor aging and/or contamination, resulting in changes in the latent features of the data collected by the sensor. For the algorithms that do not use standard samples for recalibration like the proposed method, when this change occurs, the algorithms cannot capture and model this change, and the recognition accuracy will be affected.

## 4. Conclusions

In this paper, we propose a target-domain-free domain adaptation convolutional neural network (TDACNN) to compensate for E-nose drift caused by ill-posed gas sensors. Drift is often nonlinear and unpredictable, bringing deterioration to real-life applications in many fields, which is a difficult problem that must be addressed. However, traditional domain adaptation methods usually require the participation of full-category target domain data, which is unrealistic in real cases. The proposed TDACNN method draws lessons from bionics and takes advantage of CNN feature generalization, building the backbone by using a sensor-specific multibranch structure and multiple classifiers to process different levels of features without involving target domain data. Multiple classifiers work comprehensively by the proposed MMD-based ensemble, which can evaluate and select the best-adapted result. In the training phase, an additive angular margin softmax loss is introduced to optimize the training, reduce the intraclass discrepancy, and enlarge the interclass distance. Experiments on preprocessed datasets and time sequence datasets under different settings demonstrate that the TDACNN outperforms several state-of-the-art methods in both long-time shift and short-time shift scenarios.